%% file: main.tex
\documentclass{article}
\usepackage{amsmath,graphicx,mlspconf}
\usepackage{cite, url}
\usepackage{amsmath,amssymb,amsfonts}
\usepackage{algorithmic}
\usepackage{graphicx}
\usepackage{textcomp}
\usepackage{xcolor}
\usepackage{booktabs,multirow}
\usepackage{adjustbox}
\usepackage{setspace}
\usepackage{tabularx}
\def\BibTeX{{\rm B\kern-.05em{\sc i\kern-.025em b}\kern-.08em
    T\kern-.1667em\lower.7ex\hbox{E}\kern-.125emX}}

\copyrightnotice{979-8-3503-7225-0/24/\$31.00 {\copyright}2024 IEEE}

\toappear{2024 IEEE International Workshop on Machine Learning for Signal Processing, Sept.\ 22--25, 2024, London, UK}

\title{Model-driven Heart Rate Estimation and Heart Murmur Detection based on Phonocardiogram}
\name{Jingping Nie, Ran Liu, Behrooz Mahasseni, Erdrin Azemi, and Vikramjit Mitra} 
\address{Apple Inc}

\begin{document}
\maketitle
\begin{abstract}
Acoustic signals are crucial for health monitoring, particularly heart sounds which provide essential data like heart rate and detect cardiac anomalies such as murmurs. This study utilizes a publicly available phonocardiogram (PCG) dataset to estimate heart rate using model-driven methods and extends the best-performing model to a multi-task learning (MTL) framework for simultaneous heart rate estimation and murmur detection.
Heart rate estimates are derived using a sliding window technique on heart sound snippets, analyzed with a combination of acoustic features (Mel spectrogram, cepstral coefficients, power spectral density, root mean square energy). Our findings indicate that a 2D convolutional neural network (\textbf{\texttt{2dCNN}}) is most effective for heart rate estimation, achieving a mean absolute error ($MAE$) of {1.312}\thinspace{bpm}. We systematically investigate the impact of different feature combinations and find that utilizing all four features yields the best results. The MTL model (\textbf{\texttt{2dCNN-MTL}}) achieves accuracy over 95\% in murmur detection, surpassing existing models, while maintaining an $MAE$ of {1.636}\thinspace{bpm} in heart rate estimation, satisfying the requirements stated by Association for the Advancement of Medical Instrumentation (AAMI).

\end{abstract}

\input{1_Introduction}
\input{3_Data}
\input{4_Cardiovsacular_Activity_Detection_Model}

\input{5_Results}
\input{6_Discussion}
\input{7_Conclusion}
\footnotesize
\bibliographystyle{IEEEbib}
\bibliography{strings, refs}

\end{document}

%% file: 1_Introduction.tex
\section{Introduction}
\label{sec:introduction}

The human heart, a symphony of rhythmic beats, encapsulates a wealth of physiological information. Heart sound signals, phonocardiogram (PCG) provide fundamental insights into heart rate (HR), cardiovascular diseases, stress assessment, and overall well-being~\cite{shyam2019ppgnet}. In an era where artificial intelligence of things (AIoT) has become an integral part of our daily life, the ability to non-invasively and accurately monitor HR and cardiovascular disease from diverse contexts empowers individuals to make informed decisions about their health~\cite{sahoo2022machine}. Emerging commercial-off-the-shelf customer-facing digital stethoscopes and research projects aim to empower users to self-monitor their cardiovascular activities from PCG~\cite{ hou2023arsteth}.

PCG produces two primary heart sounds, S1 and S2, caused by the closure of the atrioventricular and semilunar valves, respectively. Additional sounds like S3, S4, and heart murmurs can indicate cardiovascular diseases~\cite{koike2020audio}.
The complex real-world environments, coupled with ambient noise and noise introduced by body movement pose significant challenges to accurate HR estimation and heart murmur detection from PCGs. Traditional signal processing methods struggle with these noises due to strong distributional shifts and large variations across different scenarios~\cite{springer2014robust}.

\begin{figure}[!t]
    \centering
    \includegraphics[width=0.9\linewidth]{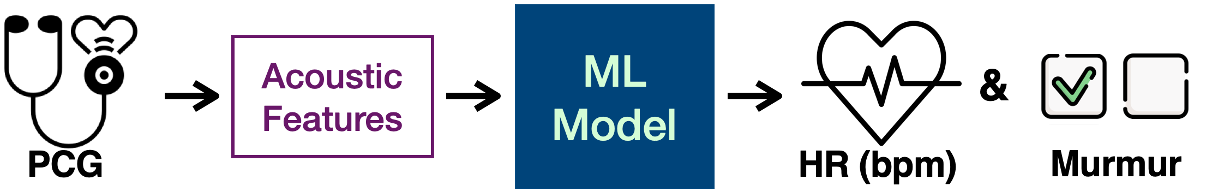}
    \vspace{-3mm}
    \caption{The overall goal of this project.}
    \label{fig:goal}
    \vspace{-\baselineskip}
\end{figure}

Deep learning, with its ability to learn intricate patterns, is emerging as a powerful tool for HR estimation, heart sound segmentation, and heart murmur detection from noisy PCGs~\cite{chorba2021deep}. Some studies use deep recurrent neural networks (RNNs) to detect S1 and S2 sounds for PCG segmentation and subsequent heart rate estimation~\cite{monteiro2022detection, fernando2019heart}. Recently, deep convolutional neural networks (CNNs) have shown superior performance in detecting abnormal heart sounds using PCGs~\cite{humayun2018learning, alkhodari2023fhsu}. 
Despite these advancements, the systematic exploration of CNN-based methods for heart rate estimation remains limited~\cite{reyna2022heart}.
This absence presents a notable challenge in the development of multi-task architectures for learning heart sounds.

Considering the aforementioned challenges and opportunities, this paper aims to investigate model-driven approaches to estimate HR and detect heart murmur from short segments of heart sounds, utilizing \emph{the CirCor DigiScope Phonocardiogram dataset}. This work first designs and compares different CNN-based models to enable reliable HR estimation and then proposes a machine learning model (\textbf{\texttt{2dCNN-MTL}}) to enable reliable HR estimation and heart murmur detection. Furthermore, the implications of this study reverberate across telemedicine, remote patient monitoring, and resource-constrained environments for better disease management, early diagnosis, and treatment. 

%% file: 3_Data.tex
\section{Data}
\label{sec:data}
\noindent\textbf{Datasets.}
The PCG dataset used in our study is the \emph{CirCor DigiScope Phonocardiogram dataset}, which includes 3,163 heart sound recordings from 942 subjects, each spanning 5.1 to 64.5 seconds and totaling about 20 hours, with half annotated, collected across four main auscultation sites in hospitals~\cite{oliveira2022circor}.
The heart sound recordings are low-pass filtered (with a cutoff frequency of $2,000\thinspace Hz$). The cardiac murmur in the dataset has been annotated in detail by an expert annotator. The segmentation annotations (onsets and offsets) regarding the location of fundamental heart sounds (S1 and S2), were obtained through a semi-supervised approach, leveraging a voting mechanism that involved three machine-learning approaches. The characteristics of PCG audio files and annotations of this dataset pose several challenges for realizing a robust HR estimation and murmur detection model. First of all, the PCG audio files are collected in the wild, where there are different low-frequency noises (e.g., environmental background noises). Second, there are annotation bias and errors in the segmentation annotations. In addition, only part of each heart sound recording is annotated.
\begin{figure}[t!]
    \centering
    \includegraphics[width=0.45\textwidth]{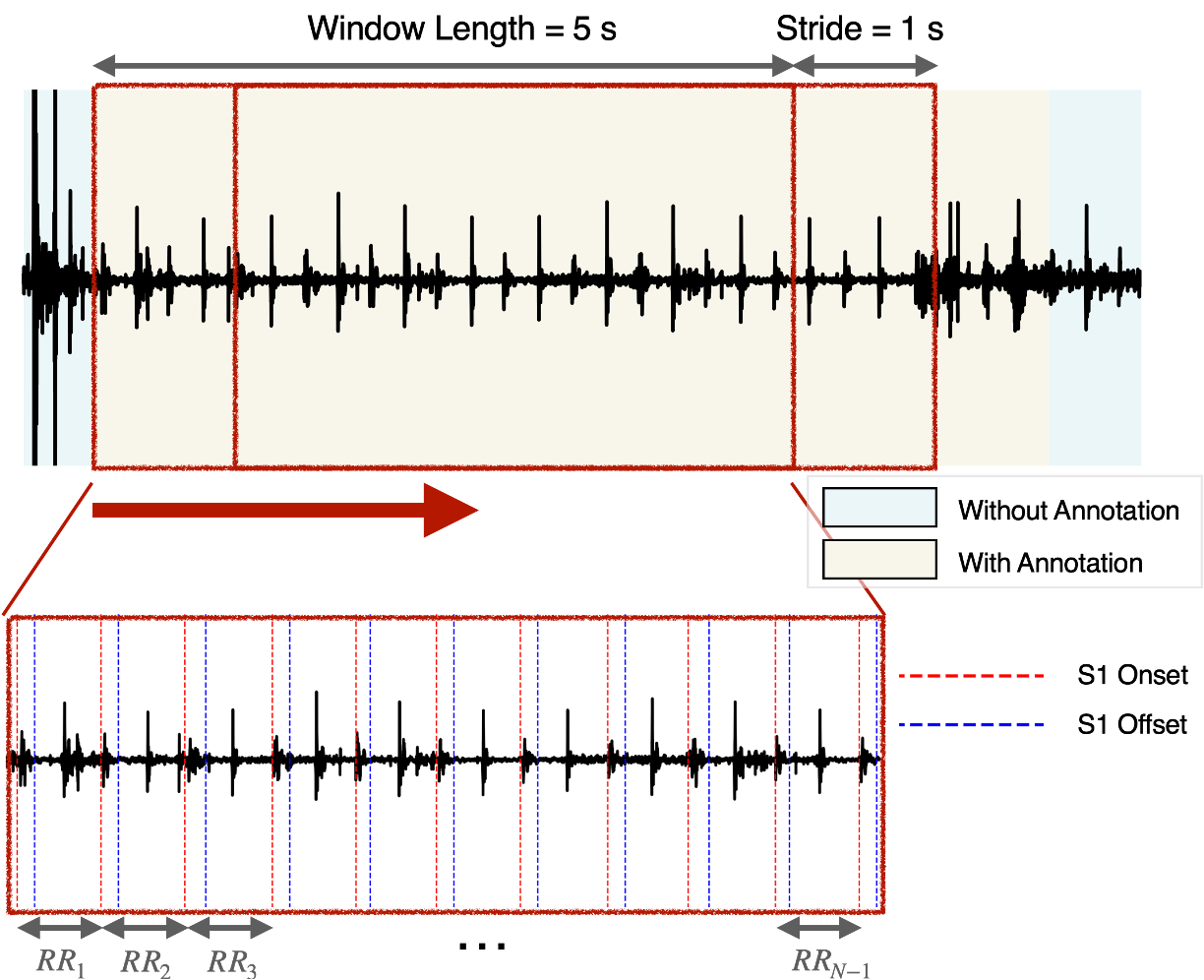}
    \vspace{-0.5\baselineskip}
    \caption{The training data preparation process.}
    \label{fig:training_preparation}
    \vspace{-0.3in}
\end{figure}

\vspace{1.0ex}
\noindent\textbf{Data Preparation.}
To prepare a dataset with a decent amount of labeled data, a sliding window with $window\_length = 5\thinspace s$ and $stride = 1\thinspace s$ is applied to the raw PCG audio files with the annotated period longer than $5\thinspace s$, as shown in Figure~\ref{fig:training_preparation}. As such, $23,381$ heart sound recording snippets are generated. The average HR ($\overline{HR}$) in beat per minute ($bpm$) of each audio snippet is calculated by $\overline{HR} = \frac{1}{N-1} \sum_{n = 1}^{N-1}\frac{60}{RR_{n}},$
where $RR_{n}$ is the interbeat interval between the adjacent onsets of S1 waves and $N$ is the number of S1 waves that appear in the audio snippet. The appearance of the heart murmur is assigned to each sound snippet ($Murmur\thinspace \in\{Absent,\thinspace Present,\thinspace Unknown\}$). The dataset is split into a training set ($80\%$), a validation set ($10\%$), and a test set ($10\%$). The audio snippets in each set are from different subjects. Note that the murmur detection only applies to the audio snippets with the murmur labels as $Absent$ or $Present$.

%% file: 4_Cardiovsacular_Activity_Detection_Model.tex
\section{Model Design and Microbenchmarks}
\label{sec:model}

In this section, we discuss the acoustic features, model design reasonings, and microbenchmarks. We start with exploring, designing, and micro benchmarking the model for the more challenging task, HR estimation, and then moving towards multi-task learning (MTL) in Section~\ref{sec:MTL model_architecture}. 

\begin{figure*}[h!]
\centering
        \includegraphics[width=0.63\textwidth]{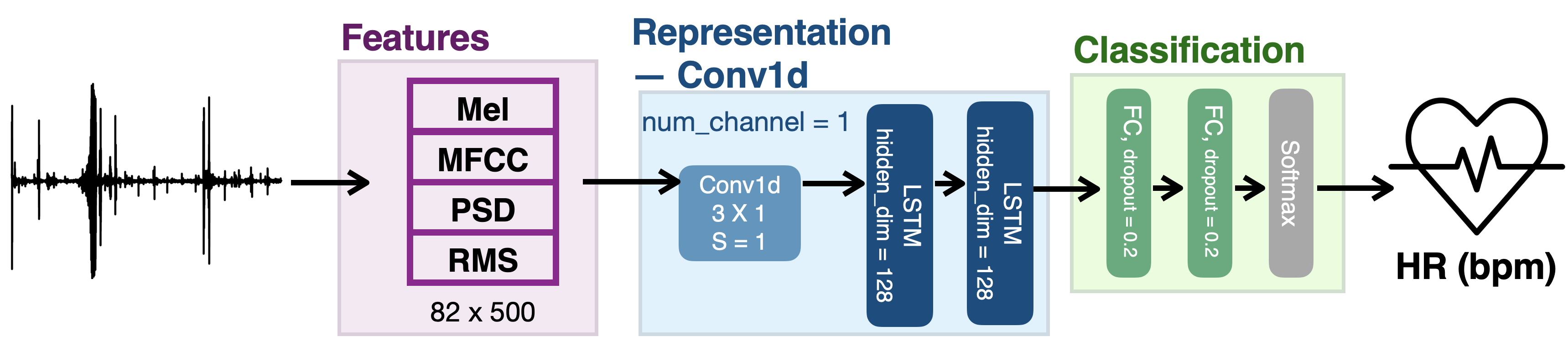}
        \caption{HR estimation based on time convolutional (1D) neural network and LSTM (\textbf{\texttt{TCNN-lstm)}}.}
        \label{fig:tcnnlstm}
        \vspace{-1\baselineskip}
\end{figure*}

\begin{figure*} [h!]
    \centering
    \includegraphics[width=0.85\textwidth]{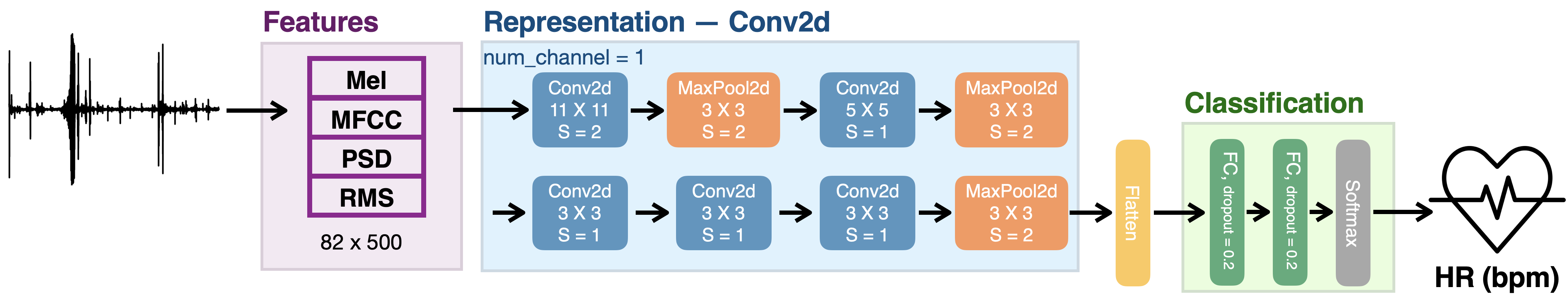}
    \vspace{-0.7\baselineskip}
    \caption{HR estimation based on 2D convolutional neural network (\textbf{\texttt{2dCNN}}).}
    \label{fig:AlexNetHR}
    \vspace{-1\baselineskip}
\end{figure*}

\subsection{Acoustic Features}
In this study, we have employed a selection of prominent acoustic features to characterize audio signals. These features contain Mel spectrogram (\texttt{Mel}), Mel-frequency cepstral coefficients (\texttt{MFCC}), power spectral density (\texttt{PSD}), and the root mean square energy (\texttt{RMS}) of the audio signal, which are well-established and widely employed in the field of audio signal processing~\cite{mitra2023investigating, xia2021csafe}. \texttt{PSD} and \texttt{RMS} contain the temporal information of each sound snippet, while \texttt{Mel} and \texttt{MFCC} provide insights for both temporal and spatial information.

To generate the acoustic features in this study, the audio is resampled from $22, 050\thinspace Hz$ to $16, 000\thinspace Hz$. For \texttt{Mel} and \texttt{MFCC}, the number of Mel bands and MFCCs is set to $40$, the highest frequency is set to $2, 000\thinspace Hz$, the window size for the short-time Fourier transform (STFT) is set to be $1,024$, and the hop length is set to be $160$.

\subsection{Model Training and Evaluation}
As HR is usually presented in an integer format and usually ranges from $40$ to $180 \thinspace bpm$~\cite{pursche2012video}. As such, we treated the HR estimation as a 141-class classification problem and the murmur detection as a binary classification task. The weighted cross entropy ($CE$) loss and binary cross-entropy ($BCE$) are used for HR (\texttt{HR}) estimation and heart murmur (\texttt{MM}) detection tasks, respectively:
\begin{eqnarray}
    &CE_{\texttt{HR}} = w_{\texttt{HR}}\sum_{a=1}^{A}(-\sum_{c=1}^{C}\log\frac{\exp{(x_{a,c})}}{\sum_{i = 1}^{C}\exp{(x_{n,i})}}y_{a,c})& \\
    &BCE_{\texttt{MM}} = w_{\texttt{MM}}\sum_{b = 1}^{B}[-y_b\dot\log x_b+(1-y_b)\dot\log(1-x_b)]& \\
    &\mathcal{L} = CE_{\texttt{HR}}+BCE_{\texttt{MM}},& ~\label{eq:multi-task}
\end{eqnarray}
where $x$ is the input, $y$ is the target, $A$ is the number of audio snippets, $B$ is the number of audio snippets containing heart murmur labels ($y_b = 0\thinspace (Absent)$ or $y_b = 1\thinspace (Present)$), $C$ is the number of HR estimation classes, $\mathcal{L}$ is the training objective for MTL, and $w_{\texttt{HR}}$ and $w_{\texttt{MM}}$ are the weights as hyperparameters. The models were trained with a mini-batch size of $16$ for $100$ epochs. The initial learning rate was $0.001$ for all models with Adam as the optimizer.

In addition, mean absolute error ($MAE_{\texttt{HR}}$) was used to evaluate the model performance for the HR estimation, while accuracy ($ACC_{\texttt{MM}}$), precision ($Precision_{\texttt{MM}}$), and recall ($Recall_{\texttt{MM}}$) were used to check the performance of murmur detection tasks. $MAE_{\texttt{HR}}$ and $ACC_{\texttt{MM}}$ are defined as:
\begin{eqnarray}
    &MAE_{\texttt{HR}} = \frac{1}{M}\sum_{i=1}^{M}|HR_{predicted, i} - HR_{target, i}|&\\
    &ACC_{\texttt{MM}} = \frac{1}{M}\sum_{i=1}^{M}\mathbf{1}\{MM_{predicted, i} = MM_{target, i}\}, &
\end{eqnarray}
where $M$ is the total number of samples in the dataset. 

\subsection{HR Estimation Model Architecture}
The learning objective was $CE_{\texttt{HR}}$. $MAE_{\texttt{HR}}$ was used to evaluate HR estimation models. The acoustic features (\texttt{Mel}, \texttt{MFCC}, \texttt{PSD}, and \texttt{RMS}) were vertically concatenated to expand the spatial information (\emph{Features} module). As the acoustics features contain temporal and spatial information, inspired by~\cite{humayun2018learning} and~\cite{su2019extended}, we first investigated the temporal information by constructing a time-convolutional long short-term memory network (\textbf{\texttt{TCNN-LSTM}}) model as shown in Figure~\ref{fig:tcnnlstm}. Specifically, there is one 1D temporal convolutional layer followed by two layers of LSTM for representation extraction (\emph{Representation} module). In addition, there are two fully connected layers with dropout to reduce overfitting and a softmax layer in the \emph{Classification} module for HR estimation.

\begin{figure}[h!]
        \centering
        \includegraphics[width=0.45\textwidth]{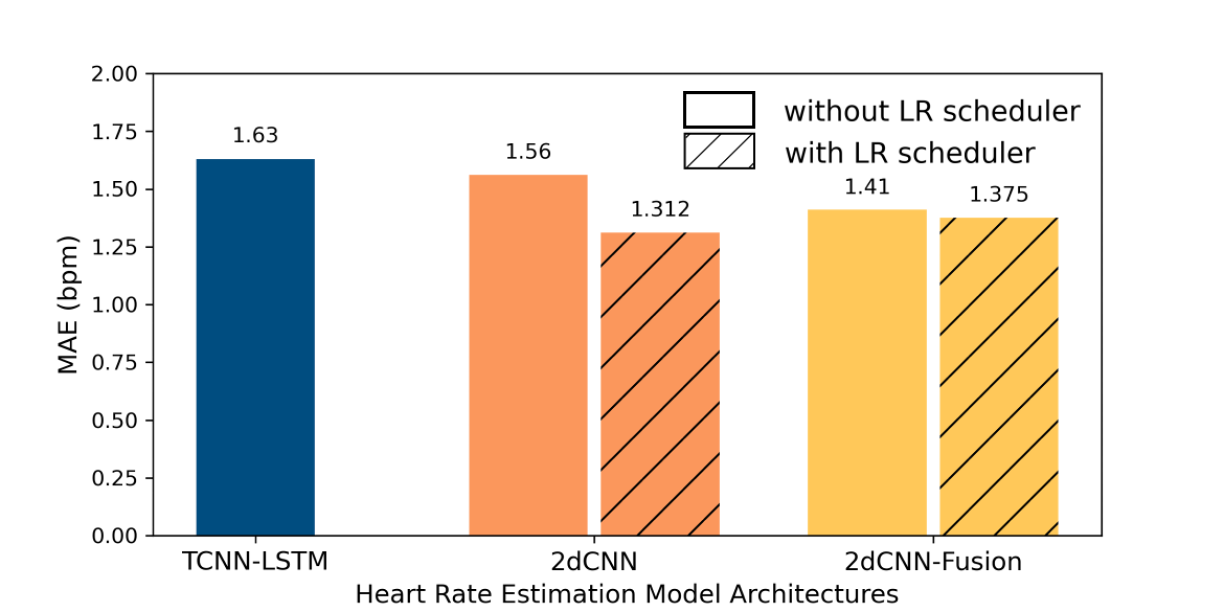}
        \vspace{-3mm}
        \caption{MAE of different HR estimation models.}
        \label{fig:HR_model_acc}
        \vspace{-\baselineskip}
\end{figure}

We then explored the information from both time- and frequency domains by designing an AlexNet-based 2D-convolutional (\textbf{\texttt{2dCNN}}) model inspired by~\cite{alafif2020normal} (see Figure~\ref{fig:AlexNetHR}). In particular, the \emph{Representation} module of this \textbf{\texttt{2dCNN}} model has five convolutional layers, each followed by max-pooling layers, where the convolutional layers use different filter sizes and strides and the activation function used is the rectified linear unit (ReLU). Then, the multi-dimensional output from the convolutional and pooling layers is transformed into a one-dimensional vector through a flattening operation before sending it to the final \emph{Classification} module. As illustrated in Figure ~\ref{fig:HR_model_acc}, the \textbf{\texttt{2dCNN}} model exhibits a smaller mean absolute error ($MAE_{\textbf{\texttt{2dCNN}}} = 1.56$) compared to the \textbf{\texttt{TCNN-LSTM}} model ($MAE_{\textbf{\texttt{TCNN-LSTM}}} = 1.63$), demonstrating a statistically significant improvement with the 2D CNN architecture.

\begin{figure}[h!]
        \centering
        \vspace{-3mm}
        \includegraphics[width=0.43\textwidth]{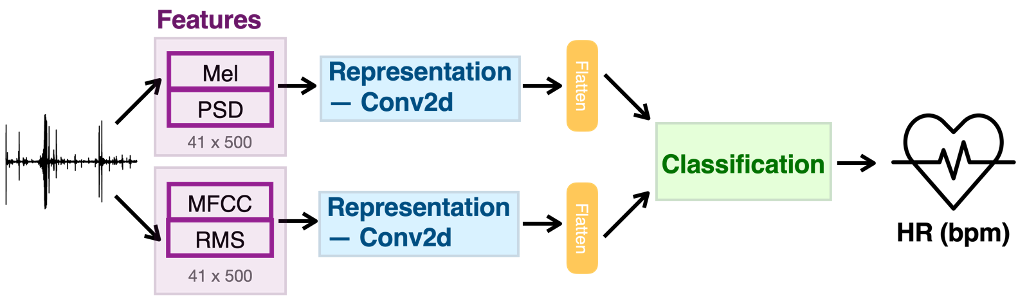}
        \caption{HR estimation based on a fusion 2D-convolutional neural network (\textbf{\texttt{2dCNN-Fusion}}).}
        \vspace{-3mm}
        \label{fig:baselineCNN}
        \vspace{-\baselineskip}
\end{figure}

In addition, to further probe if extending the temporal features would increase the model performance, a fusion 2D-convolutional model (\textbf{\texttt{2dCNN-Fusion}}) was designed as illustrated in Figure~\ref{fig:baselineCNN}. In this \textbf{\texttt{2dCNN-Fusion}} model, the vertical concatenations of \texttt{Mel} and \texttt{PSD} as well as \texttt{MFCC} and \texttt{RMS} are separately fed into two \emph{Representation} modules, each followed by flattening operations. The resultant flattened representations are then directed to the \emph{Classification} module. The \textbf{\texttt{2dCNN-Fusion}} model demonstrates a slightly improved performance ($MAE_{\textbf{\texttt{2dCNN-Fusion}}} = 1.41$) compared to the baseline \textbf{\texttt{2dCNN}} model ($MAE_{\textbf{\texttt{2dCNN}}} = 1.56$). However, it is important to highlight that the \textbf{\texttt{2dCNN}} model's performance outshines all other models, achieving the best results when a step-wise learning rate (LR) scheduling strategy is applied. This LR scheduler is activated once the validation set's $MAE$ drops below 2, with a step size set at 2 and a decay rate of 0.1. With this adaptive learning rate strategy in place, the \textbf{\texttt{2dCNN}} model yields superior results, with an even lower $MAE$ of $1.312$. Considering the performance and model size, \textbf{\texttt{2dCNN}} is selected as the model and is leveraged to enable MTL (HR estimation and murmur detection).

\subsection{Selection of Acoustic Features}
\begin{figure}[h!]
        \vspace{-\baselineskip}
        \centering
        \includegraphics[width=0.45\textwidth]{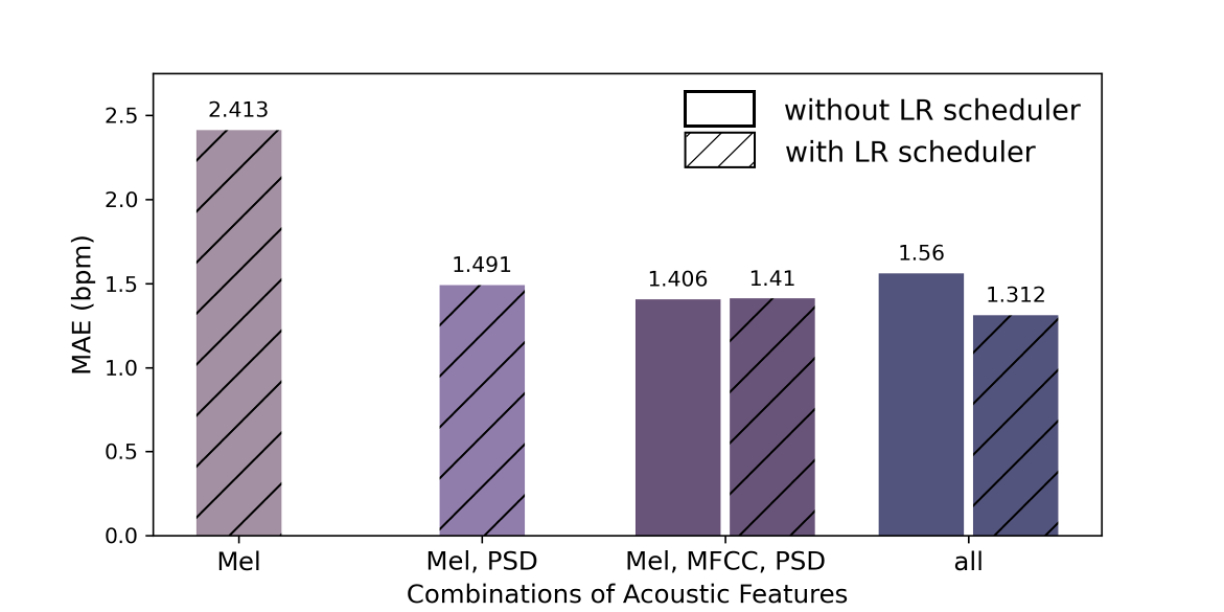}
        \vspace{-3mm}
        \caption{MAE of \textbf{\texttt{2dCNN}} model with different acoustic feature combinations.}
        \vspace{-\baselineskip}
        \label{fig:feature_selection_acc}
\end{figure}

We extensively investigated the contribution of various acoustic features to the performance of the \textbf{\texttt{2dCNN}} model. In particular, as illustrated in Figure~\ref{fig:feature_selection_acc}, we vertically concatenated different combinations of \texttt{Mel}, \texttt{MFCC}, \texttt{PSD}, and \texttt{RMS} to form the \emph{Features} module for the \textbf{\texttt{2dCNN}} model. When considering the use of solely \texttt{Mel} features within the \emph{Features} module—an approach frequently adopted in acoustics-based biosignals estimators~\cite{kumar2021estimating}—the model achieved an accuracy resulting in an $MAE_{\textbf{\texttt{2dCNN}}} = 2.413$ on the testing dataset. Although some of the other feature combinations achieve slightly better $MAE$, with the LR scheduler, using all four features leads to the best model performance.

\begin{table*}[t!]
  \centering
  \caption{The performance of \textbf{\texttt{2dCNN-MTL}} with different weights in learning objective and w/ and w/o LR scheduler.}
  \small
  \begin{tabularx}{\textwidth}{|X|X|X|X|X|X|X|}
    \hline
    \multirow{3}{*}{} & \multicolumn{2}{|c|}{$w_{\texttt{HR}} = 1, w_{\texttt{MM}}=1$, No Scheduler} & \multicolumn{2}{|c|}{$w_{\texttt{HR}} = 1, w_{\texttt{MM}}=1$, LR scheduler} & \multicolumn{2}{|c|}{$w_{\texttt{HR}} = 1, w_{\texttt{MM}}=2$, LR scheduler} \\
    \cline{2-7}
    & \small Best Model\(_{\textrm{HR}}\) & \small Best Model\(_{\textrm{MM}}\) & \small Best Model\(_{\textrm{HR}}\) & \small Best Model\(_{\textrm{MM}}\) & \small Best Model\(_{\textrm{HR}}\) & \small Best Model\(_{\textrm{MM}}\) \\
    \hline
    \small $MAE_{\textrm{HR}}$ & 1.542 & 1.636 & 1.338 & 1.368 & 1.400 & 1.455 \\
    \hline
    \small $ACC_{\textrm{MM}}$ & 92.39\% & 97.49\% & 95.19\% & 95.63\% & 95.93\% & 97.33\% \\
    \hline
    \small $Precision_{\textrm{MM}}$ & 55.11\% & 85.16\% & 68.34\% & 71.6\% & 68.20\% & 84.28\% \\
    \hline
    \small $Recall_{\textrm{MM}}$ & 81.46\% & 86.82\% & 86.34\% & 84.88\% & 86.82\% & 86.82\% \\
    \hline
  \end{tabularx}
  \label{table:MLT_ACC}
  \vspace{-\baselineskip}
\end{table*}

\section{MTL Model Architecture and Training}
\label{sec:MTL model_architecture}

\begin{figure} [h!]
\vspace{-0.5\baselineskip}
    \centering
    \includegraphics[width=0.45\textwidth]{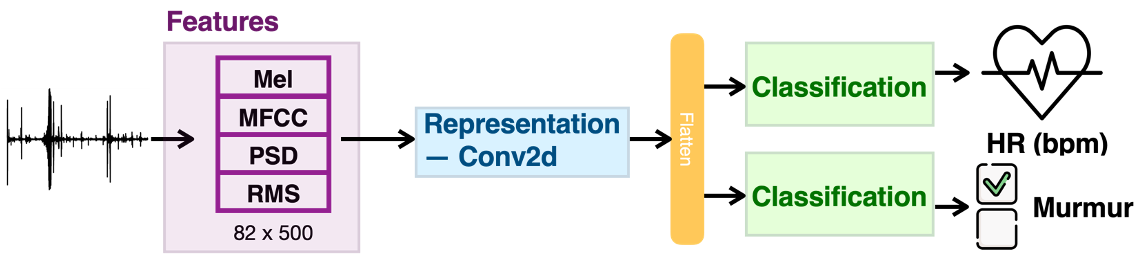}
    \vspace{-3mm}
    \caption{HR estimation and murmur detection based on 2D convolutional neural network (\textbf{\texttt{2dCNN-MTL}}).}
    \vspace{-0.5\baselineskip}
    \label{fig:AlexNet_MTL}
\end{figure}

To facilitate HR estimation and heart murmur detection within the same model, we propose the \textbf{\texttt{2dCNN-MTL}} model shown in Figure~\ref{fig:AlexNet_MTL}. This model incorporates an additional \emph{Classification} module after the flatten layer of the \textbf{\texttt{2dCNN}} model, specifically designed for murmur detection. The training objective for \textbf{\texttt{2dCNN-MTL}} model is described by Equation~\ref{eq:multi-task}. We studied different weights ($w_{\texttt{HR}}$ and $w_{\texttt{MM}}$) and the effect of adding the LR scheduler as listed in Table~\ref{table:MLT_ACC}.

Since the \textbf{\texttt{2dCNN-MTL}} model is tasked with simultaneously optimizing HR estimation and murmur detection performance, a fundamental trade-off arises when selecting the best model. 
As shown in Table~\ref{table:MLT_ACC}, when setting both $w_{\texttt{HR}}$ and $w_{\texttt{MM}}$ to 1 in the MTL loss (see Equation~\ref{eq:multi-task}), the best \textbf{\texttt{2dCNN-MTL}} models for HR estimation perform comparably to the \textbf{\texttt{2dCNN}} model, with $ACC_{\textrm{MM}}$ reaching $92.39\%$ and $95.19\%$ with and without the step-wise LR scheduler, respectively. Adding the LR scheduler significantly increases the $Precision_{\texttt{MM}}$ and $Recall_{\texttt{MM}}$. The best \textbf{\texttt{2dCNN-MTL}} models for murmur detection achieve an accuracy of $97.49\%$ but slightly decrease HR estimation performance. 
Furthermore, by increasing $w_{\texttt{MM}}$, the weights for $BCE_{\textrm{MM}}$ in the MTL loss from 1 to 2, we observed an improvement in the model performance. Specifically, the accuracy ($ACC_{\texttt{MM}}$) of the best \textbf{\texttt{2dCNN-MTL}} models increased by 0.74\% in heart rate estimation and by 1.7\% in murmur detection. Additionally, there were increases in both the precision ($Precision_{\texttt{MM}}$) and recall ($Recall_{\texttt{MM}}$) metrics.
However, this increase in $w_{\texttt{MM}}$ results in a marginal reduction in $MAE_{\textrm{HR}}$.

Association for the Advancement of Medical Instrumentation (AAMI) states that ``HR monitors should be able to compute the HR to within 10\% of the reference HR, or within five beats per minute (bpm), whichever is larger''~\cite{ANSIWebsite}. Our \textbf{\texttt{2dCNN-MTL}} model satisfies the AAMI requirements to proceed with HR estimation on the 5-second PCG snippets and surpasses the performance of the HR estimator for PCG recordings~\cite{springer2014robust}. Furthermore, across all MTL weight configurations, the murmur detection accuracy of the \textbf{\texttt{2dCNN-MTL}} model is comparable to other murmur detection models on the same dataset~\cite{lu2022lightweight, reyna2022heart}. It is worth noting that our murmur detection accuracy is computed using audio snippets with labels indicating the presence (1) or absence (0) of murmurs. In contrast, the accuracy figures reported by other projects primarily rely on the detection outcomes derived from complete audio recordings within the hidden test set.

%% file: 5_Results.tex
\section{Discussion}
\label{sec:results}

\begin{figure}
    \centering
    \includegraphics[width=0.65\linewidth]{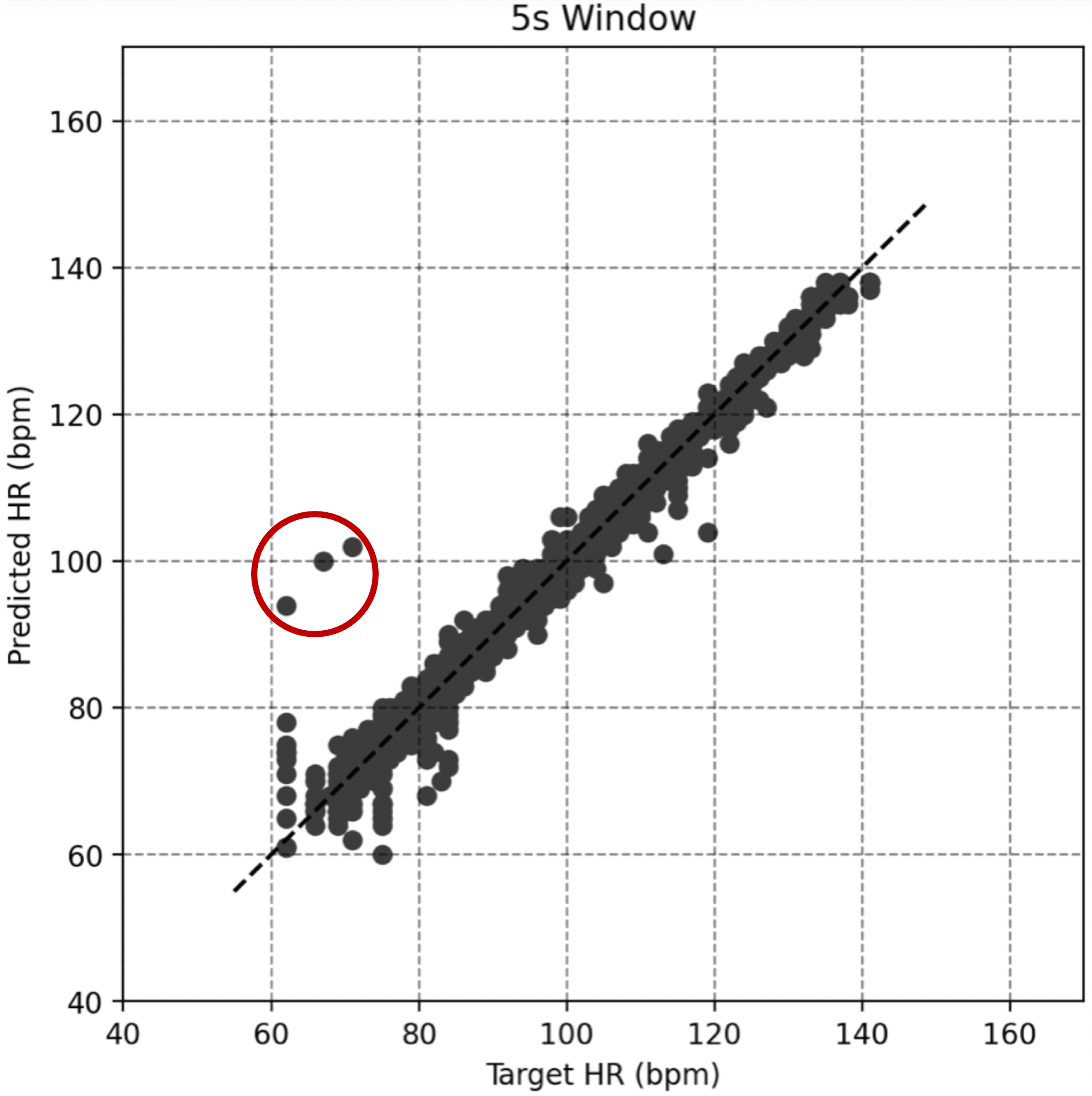}
    \vspace{-3mm}
    \caption{The predicted HR versus the target HR for the test set with \textbf{\texttt{2dCNN}} model.}
    \label{fig:res}
    \vspace{-\baselineskip}
\end{figure}

\noindent\textbf{HR Estimation.}
Figure~\ref{fig:res} shows the predicted HR alongside the target HR, using the \textbf{\texttt{2dCNN}} model that yielded the best performance on the heart sound snippets within the test set ($MAE = 1.312$). In general, the predicted results exhibit a strong correlation with the target HR, while the \textbf{\texttt{2dCNN}} model exhibits lower accuracy in estimating the HR for sound snippets with lower target HR. Notably, the outliers marked by the red circle in Figure~\ref{fig:res} pertain to sound snippets from a participant with arrhythmia (an irregular heartbeat). These snippets have interbeat intervals ($RR$) ranging from $0.68\thinspace ms$ to $1.09\thinspace ms$. Note that the annotation for the $CirCor$ dataset lacks information regarding whether the participants have concurrent heart diseases alongside heart murmurs.

In addition, we deployed a baseline signal processing-based approach to HR detection using \texttt{PSD}. This involved processing the \texttt{PSD} with a Butterworth low-pass filter at a cutoff frequency of $6\thinspace Hz$ to isolate the HR frequency band. Peak detection was performed on the filtered PSD signal to identify significant peaks corresponding to S1 and S2 waves. Peak detection was then applied to the filtered \texttt{PSD} to identify significant S1 and S2 peaks, ensuring a minimum distance of 8 samples between peaks to exclude spurious detections. We calculated inter-peak intervals for S1 and S2, converting these to beats per minute ($bpm$) to estimate the HR. Using this method, the mean absolute error is $8.2757 \thinspace bpm$, and this method can hardly estimate the sound snippets with low and high HR correctly.

\begin{figure}[!t]
    \centering
    \vspace{-3mm}
    \includegraphics[width=0.95\linewidth]{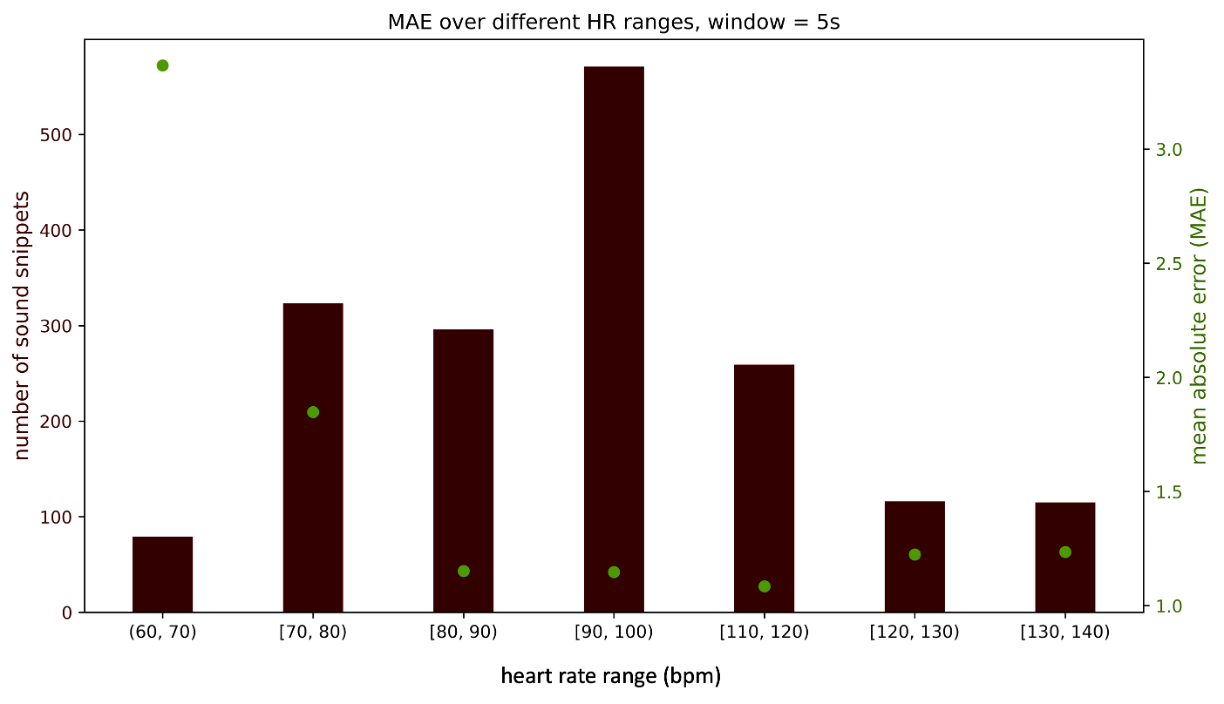}
    \vspace{-3mm}
    \caption{The \textbf{\texttt{2dCNN}} model performance on different HR ranges and the number of sound snippets in each HR range in the test set.}
    \label{fig:skewness}
    \vspace{-\baselineskip}
\end{figure}

\vspace{1.0ex}
\noindent\textbf{Data Skewness.}
To further investigate the performance of the \textbf{\texttt{2dCNN}} model across various HR ranges, we exam the distribution of sound snippets within these ranges. The histogram depicted in Figure~\ref{fig:skewness} illustrates the distribution of sound snippets across different HR ranges in the test set, revealing a degree of data skewness. In particular, the counts of sound snippets are not evenly distributed among the various HR ranges, with a notable concentration of sound snippets falling within the 90 to 100 bpm HR range. Furthermore, the $MAE$ of the predicted HR for sound snippets within each HR range is depicted as green scatter points in Figure~\ref{fig:skewness}. The \textbf{\texttt{2dCNN}} model demonstrates effective performance with sound snippets having a target HR exceeding 70 bpm, yielding an average $MAE$ of less than 2 bpm. However, its performance is less reliable when the target HR falls below 70 bpm. The uneven data distribution among HR ranges is a potential contributing factor to the model's diminished performance at lower or higher target HR. Interestingly, despite having fewer snippets in the 120--140 bpm HR range compared to the 60-80 bpm range, the model exhibits more errors for the latter. This phenomenon may be linked to the scarcity of S1 and S2 segments within the 5-second window for lower HR.

\vspace{1.0ex}
\noindent\textbf{TCNN-based MLT.}
We also deployed a \textbf{\texttt{TCNN-LSTM-MLT}} model follow the similar design of \textbf{\texttt{2dCNN-MLT}} model shown in Figure~\ref{fig:AlexNet_MTL}, which the \textit{Representation 2D} module replaced by the \textit{Representation 1D} module depicted in Figure~\ref{fig:tcnnlstm}. With $w_{\texttt{HR}} = 1$ and $w_{\texttt{MM}} = 1$, for the best HR model, \textbf{\texttt{TCNN-LSTM-MLT}} achieves $MAE_{\textrm{HR}}$ of $1.9138\thinspace bpm$ and $2.0729\thinspace bpm$, and $ACC_{\textrm{MM}}$ of 90.14\% and 89.7\% with and without the LR scheduler. \textbf{\texttt{2dCNN-MLT}} outperforms \textbf{\texttt{TCNN-LSTM-MLT}} in the multi-task learning setup.

%% file: 6_Discussion.tex
\section{Limitation and Future Work}\label{sec:discussion}
This section summarizes current limitations and areas for further exploration. The \emph{CirCor} dataset lacks annotations for environmental noise and respiration rate and contains low-pass filtered PCG audio files, so our method does not include explicit source separation or noise suppression steps. It would be beneficial for future studies to investigate and incorporate heart sound source separation method to remove low-frequency noises without losing acoustic features for heart murmurs. Additionally, 
considering the duration of PCG files in the \emph{CirCor} dataset, as detailed in Section~\ref{sec}, we set the window and stride lengths to $5\thinspace s$ and $1\thinspace s$, respectively,  to generate an adequate number of heart sound snippets for training. However, reducing the window size to 3s with the same \textbf{\texttt{2dCNN-MTL}} model increases the MAE for HR estimation to $3.295\thinspace bpm$. We also plan to implement a custom loss function with a penalty term weighted by the difference between predicted and true heart rates to ensure larger errors are penalized more heavily. In current model settings, treating heart rate estimation as a regression problem has underperformed compared to treating it as a classification problem. We aim to explore more regression models and perform hyperparameter tuning to investigate the feasibility of using regression for heart rate estimation. Furthermore, it is worth noting that the PCG recordings in the \emph{CirCor} dataset are resting heart sounds. The exploration of non-steady-state PCG data, such as post-exercise heart sounds, could significantly enhance model adaptability across various everyday scenarios and enable more applications.

%% file: 7_Conclusion.tex
\section{Conclusion}\label{sec:conclusion}
This study presents a significant contribution to the field of health monitoring and cardiac assessment through its novel model-driven approach to heart rate estimation and heart murmur detection based on phonocardiogram (PCG) analysis. Utilizing a publicly available PCG dataset, the research demonstrated the efficacy of the proposed 2D convolutional neural network (\textbf{\texttt{2dCNN}}) for heart rate estimation. The model, with a mean absolute error ($MAE$) of $1.312\thinspace bpm$, effectively integrates diverse acoustic features: \texttt{Mel}, \texttt{MFCC}, \texttt{PSD}, and \texttt{RMS}. This work extended to a multi-task learning (MTL) framework, encapsulated in the \textbf{\texttt{2dCNN-MTL}} model, which concurrently achieved heart rate estimation and murmur detection. The \textbf{\texttt{2dCNN-MTL}} model's accuracy exceeds 95\%, surpassing existing models in both accuracy and efficiency, with a maintained $MAE$ below $1.636\thinspace bpm$ in heart rate estimation. We envision the integration of these techniques to revolutionize remote patient monitoring and self-care.

%% file: main.bbl
\begin{thebibliography}{10}

\bibitem{shyam2019ppgnet}
A~Shyam, Vignesh Ravichandran, SP~Preejith, Jayaraj Joseph, and Mohanasankar Sivaprakasam,
\newblock ``Ppgnet: Deep network for device independent heart rate estimation from photoplethysmogram,''
\newblock in {\em 2019 41st Annual International Conference of the IEEE Engineering in Medicine and Biology Society (EMBC)}. IEEE, 2019, pp. 1899--1902.

\bibitem{sahoo2022machine}
Goutam~Kumar Sahoo, Keerthana Kanike, Santos~Kumar Das, and Poonam Singh,
\newblock ``Machine learning-based heart disease prediction: A study for home personalized care,''
\newblock in {\em 2022 IEEE 32nd International Workshop on Machine Learning for Signal Processing (MLSP)}. IEEE, 2022, pp. 01--06.

\bibitem{hou2023arsteth}
Kaiyuan Hou, Stephen Xia, Emily Bejerano, Junyi Wu, and Xiaofan Jiang,
\newblock ``{ARSteth}: Enabling home self-screening with ar-assisted intelligent stethoscopes,''
\newblock in {\em Proceedings of the 22nd International Conference on Information Processing in Sensor Networks}, 2023, pp. 205--218.

\bibitem{koike2020audio}
Tomoya Koike, Kun Qian, Qiuqiang Kong, Mark~D Plumbley, Bj{\"o}rn~W Schuller, and Yoshiharu Yamamoto,
\newblock ``Audio for audio is better? an investigation on transfer learning models for heart sound classification,''
\newblock in {\em 2020 42nd Annual International Conference of the IEEE Engineering in Medicine \& Biology Society (EMBC)}. IEEE, 2020, pp. 74--77.

\bibitem{springer2014robust}
David~B Springer, Thomas Brennan, Jens Hitzeroth, Bongani~M Mayosi, Lionel Tarassenko, and Gari~D Clifford,
\newblock ``Robust heart rate estimation from noisy phonocardiograms,''
\newblock in {\em Computing in Cardiology 2014}. IEEE, 2014, pp. 613--616.

\bibitem{chorba2021deep}
John~S Chorba, Avi~M Shapiro, Le~Le, John Maidens, John Prince, Steve Pham, Mia~M Kanzawa, Daniel~N Barbosa, Caroline Currie, Catherine Brooks, et~al.,
\newblock ``Deep learning algorithm for automated cardiac murmur detection via a digital stethoscope platform,''
\newblock {\em Journal of the American Heart Association}, vol. 10, no. 9, pp. e019905, 2021.

\bibitem{monteiro2022detection}
Sofia Monteiro, Ana Fred, and Hugo~Pl{\'a}cido da~Silva,
\newblock ``Detection of heart sound murmurs and clinical outcome with bidirectional long short-term memory networks,''
\newblock in {\em 2022 Computing in Cardiology (CinC)}. IEEE, 2022, vol. 498, pp. 1--4.

\bibitem{fernando2019heart}
Tharindu Fernando, Houman Ghaemmaghami, Simon Denman, Sridha Sridharan, Nayyar Hussain, and Clinton Fookes,
\newblock ``Heart sound segmentation using bidirectional lstms with attention,''
\newblock {\em IEEE journal of biomedical and health informatics}, vol. 24, no. 6, pp. 1601--1609, 2019.

\bibitem{humayun2018learning}
Ahmed~Imtiaz Humayun, Shabnam Ghaffarzadegan, Zhe Feng, and Taufiq Hasan,
\newblock ``Learning front-end filter-bank parameters using convolutional neural networks for abnormal heart sound detection,''
\newblock in {\em 2018 40th Annual International Conference of the IEEE Engineering in Medicine and Biology Society (EMBC)}. IEEE, 2018, pp. 1408--1411.

\bibitem{alkhodari2023fhsu}
Mohanad Alkhodari, Murad Almadani, Samit~Kumar Ghosh, and Ahsan~H Khandoker,
\newblock ``Fhsu-net: Deep learning-based model for the extraction of fetal heart sounds in abdominal phonocardiography,''
\newblock in {\em 2023 IEEE 33rd International Workshop on Machine Learning for Signal Processing (MLSP)}. IEEE, 2023, pp. 1--6.

\bibitem{reyna2022heart}
Matthew~A Reyna, Yashar Kiarashi, Andoni Elola, Jorge Oliveira, Francesco Renna, Annie Gu, Erick A~Perez Alday, Nadi Sadr, Ashish Sharma, Sandra Mattos, et~al.,
\newblock ``Heart murmur detection from phonocardiogram recordings: The george b. moody physionet challenge 2022,''
\newblock in {\em 2022 Computing in Cardiology (CinC)}. IEEE, 2022, vol. 498, pp. 1--4.

\bibitem{oliveira2022circor}
Jorge Oliveira, Francesco Renna, Paulo Costa, Marcelo Nogueira, Ana~Cristina Oliveira, Andoni Elola, Carlos Ferreira, Alipio Jorge, Ali~Bahrami Rad, Matthew Reyna, et~al.,
\newblock ``The circor digiscope phonocardiogram dataset,''
\newblock {\em version 1.0. 0}, 2022.

\bibitem{mitra2023investigating}
Vikramjit Mitra, Jingping Nie, and Erdrin Azemi,
\newblock ``Investigating salient representations and label variance in dimensional speech emotion analysis,''
\newblock {\em arXiv preprint arXiv:2312.16180}, 2023.

\bibitem{xia2021csafe}
Stephen Xia, Jingping Nie, and Xiaofan Jiang,
\newblock ``Csafe: An intelligent audio wearable platform for improving construction worker safety in urban environments,''
\newblock in {\em Proceedings of the 20th International Conference on Information Processing in Sensor Networks}, 2021, pp. 207--221.

\bibitem{pursche2012video}
Thomas Pursche, Jarek Krajewski, and Reinhard Moeller,
\newblock ``Video-based heart rate measurement from human faces,''
\newblock in {\em 2012 IEEE international conference on consumer electronics (ICCE)}. IEEE, 2012, pp. 544--545.

\bibitem{su2019extended}
Yuanhang Su and C-C~Jay Kuo,
\newblock ``On extended long short-term memory and dependent bidirectional recurrent neural network,''
\newblock {\em Neurocomputing}, vol. 356, pp. 151--161, 2019.

\bibitem{alafif2020normal}
Tarik Alafif, Mehrez Boulares, Ahmed Barnawi, Talal Alafif, Hassan Althobaiti, and Ali Alferaidi,
\newblock ``Normal and abnormal heart rates recognition using transfer learning,''
\newblock in {\em 2020 12th International Conference on Knowledge and Systems Engineering (KSE)}. IEEE, 2020, pp. 275--280.

\bibitem{kumar2021estimating}
Agni Kumar, Vikramjit Mitra, Carolyn Oliver, Adeeti Ullal, Matt Biddulph, and Irida Mance,
\newblock ``Estimating respiratory rate from breath audio obtained through wearable microphones,''
\newblock in {\em 2021 43rd Annual International Conference of the IEEE Engineering in Medicine \& Biology Society (EMBC)}. IEEE, 2021, pp. 7310--7315.

\bibitem{ANSIWebsite}
{ANSI/AAMI/IEC},
\newblock ``Ansi/aami ec13-2002: Cardiac monitors, heart rate meters, and alarms,'' 2002.

\bibitem{lu2022lightweight}
Hui Lu, Julia~Beatriz Yip, Tobias Steigleder, Stefan Grie{\ss}hammer, Maria Heckel, Naga Venkata Sai~Jitin Jami, Bjoern Eskofier, Christoph Ostgathe, and Alexander Koelpin,
\newblock ``A lightweight robust approach for automatic heart murmurs and clinical outcomes classification from phonocardiogram recordings,''
\newblock in {\em 2022 Computing in Cardiology (CinC)}. IEEE, 2022, vol. 498, pp. 1--4.

\end{thebibliography}
